\documentstyle[12pt]{article}
\def\bm#1{{\mbox{\boldmath $#1$}}}
\def\frac#1#2{ {{#1} \over {#2} }}
\def\rat#1#2{\mbox{\small $\frac{#1}{#2}$}}
\def\sub#1#2{{#1}_{\mbox{\scriptsize{#2}}}}

\def\VEV#1{\left\langle #1\right\rangle}

\def\beq{\begin{equation}}
\def\beeq{\begin{eqnarray}}
\def\eeq{\end{equation}}
\def\eeeq{\end{eqnarray}}

\def\as{\sub\alpha{S}}
\def\asZ{\as(\sub M Z)}
\def\cN{{\cal N}}
\def\ee{e$^+$e$^-$}
\def\half{\rat 1 2}
\def\ho#1{{\cal O}(\as^{#1})}

\def\nlf{N_f}
\def\oas{{\cal O}(\as)}

\def\rs{\sqrt{s}}
\def\st{\sub\sigma{tot}}
\def\tdt{t\frac{\partial}{\partial t}}

\def\Z0{Z$^0$}
\def\HW{{\small HERWIG}}
\def\JS{{\small JETSET}}
\def\pl#1#2#3{
        {\it Phys.\ Lett.\ }{\bf #1} (19#3) #2}

\def\zp#1#2#3{
        {\it Zeit.\ Phys.\ }{\bf #1} (19#3) #2}

\def\prl#1#2#3{
        {\it Phys.\ Rev.\ Lett.\ }{\bf #1} (19#3) #2}

\def\prep#1#2#3{
        {\it Phys.\ Rep.\ }{\bf #1} (19#3) #2}
\def\pr#1#2#3{
        {\it Phys.\ Rev.\ }{\bf #1} (19#3) #2}
\def\np#1#2#3{
        {\it Nucl.\ Phys.\ }{\bf #1} (19#3) #2}

\def\JETP#1#2#3{
        {\it Sov.\ Phys.\ JETP} {\bf #1} (19#3) #2}
\def\Jlet#1#2#3{
        {\it JETP Lett.\ } {\bf #1} (19#3) #2}
\def\sj#1#2#3{
        {\it Sov.\ J.\ Nucl.\ Phys.\ } {\bf #1} (19#3) #2}
\def\ar#1#2#3{
        {\it Ann.\ Rev.\ Nucl.\ Part.\ Sci.\ } {\bf #1} (19#3) #2}

\def\cern#1{CERN preprint TH.#1}
\def\cpc#1#2#3{{\em Computer Phys.\ Comm.\ }{\bf #1} (19#3) #2}

\begin{document}
\begin{titlepage}
\renewcommand{\thefootnote}{\fnsymbol{footnote}}
\begin{flushright}
     Cavendish--HEP--94/17 \\
     hep-ph/9411384 \\
\end{flushright}
\vspace*{\fill}
\begin{center}
{\Large \bf
HADRONIZATION\footnote{Lectures at Summer School on Hadronic
Aspects of Collider Physics, Zuoz, Switzerland, August 1994.}}
\end{center}
\par \vskip 2mm
\begin{center}
        {\bf B.R.\ Webber} \\
        Cavendish Laboratory, University of Cambridge,\\
        Madingley Road, Cambridge CB3 0HE, U.K.
\end{center}
\par \vskip 2mm
\begin{center} {\large \bf Abstract} \end{center}
\begin{quote}
Hadronization corrections to the predictions of perturbative QCD
are reviewed.  The existing models for the conversion of quarks
and gluons into hadrons are summarized.  The most successful
models give a good description of the data on \ee\ event shapes
and jet fragmentation functions, and suggest that the dominant
hadronization effects have a $1/Q$ dependence on the hard process
energy scale $Q$. In several cases the $1/Q$ terms can be understood
in terms of a simple longitudinal phase-space model. They can also
be inferred by relating non-perturbative renormalon effects to the
infrared cutoff dependence of perturbative contributions.
\end{quote}
\vspace*{\fill}
\begin{flushleft}
     Cavendish--HEP--94/17\\
     hep-ph/9411384 \\
     November 1994
\end{flushleft}
\end{titlepage}
\renewcommand{\thefootnote}{\fnsymbol{footnote}}

\section{Introduction}
Hadronic jet production at high energy colliders has proved
to be one of the most valuable testing-grounds for quantum
chromodynamics (QCD).  At high energies, or more precisely
at large momentum transfers, the QCD coupling $\as$
becomes small and perturbation theory becomes more
reliable. Perturbative predictions to next-to-leading
order, and to higher order in a few cases, give a good
account of the broad features of jet production
processes in \ee\, hadron-hadron, and lepton-hadron
collisions [\ref{QCD20},\ref{WebGla}].

A serious barrier to further progress in jet physics,
however, is our lack of understanding of the process
of hadronization, in which the quarks and gluons
of perturbative QCD are converted into the hadrons
that are seen in the detectors.  On general grounds,
we expect that hadronization and other non-perturbative
effects should give rise to power-suppressed
corrections to quantities that are computable
in perturbation theory. By power-suppressed, we mean
corrections proportional to $1/Q^p$, where $Q$ is
the hard process scale (the centre-of-mass energy
in \ee\ annihilation). At present there are no
solid arguments to exclude contributions with $p<2$
for observables like \ee\ event shapes, which are
not fully inclusive with respect to final-state hadrons.
Indeed, as we shall see, there are strong indications
that the leading corrections are proportional to $1/Q$
in these cases.
This is in contrast to deep inelastic structure functions
and the total \ee\ hadronic cross section, where arguments
based on the operator product expansion suggest that the
dominant power corrections should decrease like $1/Q^2$ and
$1/Q^4$ respectively [\ref{mueller}].

\begin{figure}[htb]
\vspace{6.0cm}
\includegraphics{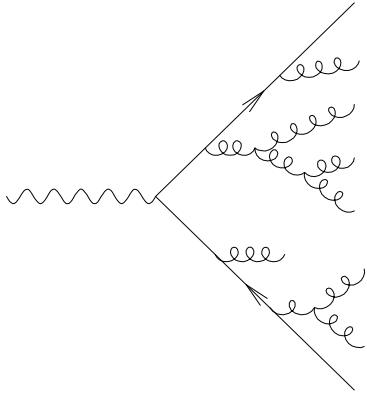}
  \caption[DATA]{
Parton cascade in \ee\ annihilation.
     }\label{fig_eeps}
\end{figure}

The general picture of jet production and fragmentation which
has developed and proved highly successful over the past decade
has three fairly distinct stages [\ref{bwarns}]. First, on a scale of
energy and time characterized by a large momentum transfer-squared
$Q^2$, a hard subprocess takes place involving a small number of primary
partons (quarks and/or gluons). For example, in \ee\ annihilation
the hard subprocess would be \ee\ $\to q\bar q$ or, more rarely,
\ee\ $\to q\bar q g$.  Next, over a period characterized by
scales $t$ such that $Q^2>t>t_0$, the primary partons develop
into multi-parton cascades or showers by multiple gluon bremsstrahlung
(Fig.~\ref{fig_eeps}). These cascades, which tend to develop along
the directions of the primary partons owing to the collinear
enhancements in QCD matrix elements, are the precursors of the
jets that are observed experimentally. The showering cutoff
scale, $t_0$, should be much greater than the intrinsic QCD
scale $\Lambda^2$, but is otherwise somewhat arbitrary, being
set by the requirement that a perturbative description
in terms of partons should remain appropriate down to
that scale.

After the parton shower has
terminated, we are left with a set of partons with virtualities
(virtual masses-squared) of the order of the cutoff scale $t_0$.
{}From this point we enter the low momentum-transfer, long-distance
regime in which non-perturbative effects cannot be neglected.
The most important of these is hadronization, which converts
the partons into the observed hadrons. As long as this stage
of the process involves only small momentum transfers, presumably
on a scale set by the QCD scale $\Lambda\sim 250$ MeV, it can
be expected to lead to power corrections to quantities that
are finite in perturbation theory, which are determined mainly
by the two earlier stages of jet production and development.

At present the only detailed descriptions of the hadronization
process are provided by models, which are discussed briefly in
the following section.  In Sect.~\ref{sec_evt}, we consider
the predictions of these models for hadronization corrections
in \ee\ event shapes, where a rather clear pattern of $1/Q$
corrections emerges.

Next, in Sect.~\ref{sec_frag}, we turn to
a discussion of hadron energy spectra in jet fragmentation.
Here hadronization enters in two distinct ways. First, the
the shapes and normalizations of the spectra of the various
hadronic species provide detailed tests of hadronization
models. Such quantities cannot be computed from perturbation
theory, because they involve the observation of individual
hadrons. However, once a fragmentation spectrum has been measured
at a particular hard process energy scale $Q$, then its
form at any other large scale can be predicted perturbatively
using the factorization properties of QCD. The only
ambiguity in the prediction, apart from the effect of
higher-order terms that can be computed in principle,
is again due to unknown power-suppressed corrections.
We review the evidence on these corrections, which are
poorly understood compared with the analogous (higher-twist)
terms in deep inelastic scattering.

One formal theoretical approach to power corrections
that may be useful is the study of infrared renormalons,
which are generated by the divergence of perturbation theory at
high orders.  Renormalons are associated
with power-suppressed effects but it is not so clear what they
have to do with the hadronization process.  However, it appears
likely that there is a renormalon contribution to event shapes,
corresponding to a $1/Q$ correction. This is discussed in
Sect.~\ref{sec_pow}. Finally, in Sect.~\ref{sec_conc},
some conclusions are drawn.

\section{Hadronization models}
\label{sec_mod}
One general approach to hadronization, based on the observation
that perturbation theory seems to work well down to rather
low scales, is the hypothesis of {\em local parton-hadron duality}
[\ref{LPHD}]. Here one supposes only that the flow of momentum
and quantum numbers at the hadron level tends to follow the
flow established at the parton level. Thus, for example, the
flavour of the quark initiating a jet should be found in a
hadron near the jet axis. The extent to which the hadron flow
deviates from the parton flow reflects the irreducible smearing
of order $\Lambda$ due to hadron formation.  Perhaps the most
striking example of local parton-hadron duality is the shape of the
hadron spectrum in jet fragmentation at relatively low energies,
which will be discussed in Sect.~\ref{sec_frag}.4.

The simplest more explicit hadronization model [\ref{Fey}]
is the longitudinal phase-space or `tube' model,\footnote{This model
is essentially the simplest version of the string model; we call it
a tube to avoid confusion with the more sophisticated Lund string
model discussed below.} in which a parton
(or, more realistically, a colour-connected pair of partons) produces
a jet of light hadrons which occupy a tube in $(y,p_t)$-space, where
$y=\half\log[(E+p_z)/(E-p_z)]$ is rapidity and $p_t$ is transverse
momentum, both measured with respect to the direction of the
initial parton.  If the hadron density in this space is $\rho(p_t)$,
the energy and momentum of a tube of rapidity length $Y$ are
\beeq\label{tube}
E &=& \int_0^Y dy\,d^2p_t \rho(p_t) p_t\cosh y = \lambda \sinh Y
\nonumber \\
P &=& \int_0^Y dy\,d^2p_t \rho(p_t) p_t\sinh y = \lambda (\cosh Y -1)
\sim E-\lambda\; ,
\eeeq
where $\lambda = \int d^2p_t \rho(p_t) p_t$ sets the hadronization scale.
Notice that the jet momentum $P$ receives a negative hadronization
correction of relative order $\lambda/E = 2\lambda/Q$ for a two-jet
configuration of total energy $Q$. Thus one generally expects
hadronization effects to scale with energy like $1/Q$.

{}From Eqs.~(\ref{tube}) we expect a mean-square hadronization
contribution to jet masses of
\beq
\sub{\VEV{M^2}}{had} = E^2-P^2 \sim \lambda Q\;.
\eeq
Comparing the perturbative predictions for jet masses with experiment,
one finds that a hadronization correction corresponding to
\beq\label{lamval}
\lambda\sim 0.5\;\mbox{GeV}
\eeq
is required. Note that this implies a fairly large jet
mass, about $7$ GeV at $Q\sim\sub{M}{Z}$, in addition to
the perturbative contribution.

We shall see that the above simple model successfully describes
many of the gross features of hadronization. In order to make more
detailed predictions, we need a specific model
for the mechanism of hadronization.  Over the years, three classes of
models have been developed, which we outline briefly in the following
subsections.

\subsection{Independent fragmentation model}
The simplest scheme for generating hadron distributions from those of partons
is to suppose that each parton fragments independently.  The original approach
of Field and Feynman [\ref{FF}] was designed to reproduce the limited
transverse momenta and approximate scaling of energy fraction distributions
observed in quark jets produced in \ee\ annihilation at moderate energies.
The fragmenting quark is combined with an antiquark from a $q\bar q$ pair
created out of the vacuum, to give a ``first-generation" meson with energy
fraction $z$. The leftover quark, with energy fraction $(1-z)$, is fragmented
in the same way, and so on until the leftover energy falls below some cutoff.
Scaling follows from the energy independence of the distribution assumed for
$z$, which is known as the fragmentation function.  The limited transverse
momenta  come from the relative transverse momenta of the created
$q\bar q$ pairs, which are given a Gaussian distribution.

For gluon fragmentation, the gluon is first split into a quark-antiquark pair,
either assigning all the gluon's momentum to one or the other ($z=0$ or 1)
with equal probability [\ref{HOSWZ}], so that the gluon behaves at a quark of
random flavour, or using the $g\to q\bar q$ Altarelli-Parisi splitting function
[\ref{APKW}].

With about four parameters to describe the fragmentation function, the
width of the transverse momentum distribution, the ratio of strange to
nonstrange pair creation, and the ratio of vector to pseudoscalar meson
production, the model proved quite successful in describing the broad
features of two-jet and three-jet final states in \ee\ annihilation at
moderate energies [\ref{HOSWZ}-\ref{AKKW}].

A weakness of the independent fragmentation scheme, as formulated above,
is that the fragmentation of a parton is supposed to depend on its energy
rather than its virtuality.  Indeed, the fragmenting parton is usually
assumed to remain on mass shell, leading to violations of momentum
conservation that have to be corrected by rescaling momenta after
hadronization is completed.  The residual colour and flavour of the
leftover parton in each jet also have to be neutralized at this stage.
There are further problems when jets become close together in angle.
Instead of merging smoothly together into a single jet, as would
happen if their fragmentation depended on their combined effective
mass, even two precisely collinear jets remain distinguishable from
a single jet.

\subsection{String model}
The string model of hadronization [\ref{AM}-\ref{AGIS}] is most easily
described for \ee\ annihilation.  Neglecting for the moment the
possibility of gluon bremsstrahlung, the produced quark and antiquark
move out in opposite directions, losing energy to the colour field,
which is supposed to collapse into a stringlike configuration between
them.  The string has a uniform energy per unit length, corresponding
to a linear quark confining potential, which is consistent with
quarkonium spectroscopy.  The string breaks up into hadron-sized
pieces through spontaneous $q\bar q$ pair production in its intense
colour field.

In practice, the string fragmentation approach does not look very
different from independent fragmentation for the simple quark-antiquark
system.  The string may be broken up starting at either the quark or the
antiquark end, or both simultaneously (the breaking points have spacelike
separations, so their temporal sequence is frame dependent), and it
proceeds iteratively by $q\bar q$ pair creation, as in independent
fragmentation.  What one gains is a more consistent and covariant picture,
together with some constraints on the fragmentation function [\ref{AGS}],
to ensure independence of whether one starts at the quark or the
antiquark, and on the transverse momentum distribution [\ref{AGIS}],
which is now related to the tunnelling mechanism by which $q\bar q$
pairs are created in the colour field of the string.

The string model becomes more distinct from independent fragmentation
when gluons are present [\ref{Sjo}].  These are supposed to produce
kinks on the string, each initially carrying localized energy and
momentum equal to that of its parent gluon.  The fragmentation of the
kinked string leads to an angular distribution of hadrons in \ee\
three-jet final states that is different from that predicted by
independent fragmentation and in better agreement with experiment
[\ref{Bar}].

For multiparton states, such as those produced by
parton showering at high $Q^2$ (Fig.~\ref{fig_eeps}),
there is ambiguity about how strings should be connected between the
various endpoints (quarks and antiquarks) and kinks (gluons).  However,
to leading order in $N_c^{-2}$ where $N_c=3$ is the number of colours,
it is always possible to arrange the produced partons in a {\em planar}
configuration, such that each has an equal and opposite colour to that
of a neighbour (both neighbours, in the case of a gluon), like the
quark and antiquark in the simplest \ee\ final state. The natural
prescription is then to stretch the string between colour-connected
neighbours, so as to make colour singlet strings of minimum
invariant mass. The reformulation of parton showers in terms
of sequential splitting of colour dipoles [\ref{dipole}] leads
to the same rule for string connection.

\subsection{Cluster model}
An important property of the parton showering process is the
{\em preconfinement} of colour [\ref{prec}].
Preconfinement implies that the pairs of colour-connected neighbouring
parton discussed above have an asymptotic mass distribution that falls
rapidly at high masses and is asymptotically $Q^2$-independent and
universal.  This suggests a class of cluster hadronization
models, in which colour-singlet clusters of partons form after
the perturbative phase of jet development and then decay into
the observed hadrons.

The simplest way for colour-singlet clusters to form after the
parton cascade is through non-perturbative splitting of gluons
into $q\bar q$ pairs [\ref{FW}].  Neighbouring colour-connected
quarks and antiquarks can then combine into singlet clusters.
The resulting cluster mass spectrum is again universal and steeply
falling at high masses.  Its precise form is determined by the QCD
scale $\Lambda$, the parton shower cutoff $t_0$, and to a lesser
extent the gluon-splitting mechanism. Typical cluster masses are
normally two or three times $\sqrt{t_0}$.

If a low value of the cutoff $t_0$ is used, of the order of 1 GeV$^2$ or less,
most clusters have masses of up to a few GeV/c$^2$ and it is reasonable
to treat them as superpositions of meson resonances.  In a popular
model [\ref{FW},\ref{clus}], each such cluster is assumed to decay
isotropically in its rest frame into a pair of hadrons, with branching
ratios determined simply by density of states.  The reduced phase space
for cluster decay into heavy mesons and baryons is then sufficient to
account fairly well for the multiplicities of the various
kinds of hadrons observed in \ee\ final states.  Furthermore, the
hadronic energy and transverse momentum distributions agree well
with experiment, without the introduction of any adjustable fragmentation
functions.  Also, the angular distribution in \ee\ three-jet
events is successfully described, as in the string model [\ref{Bar}],
provided soft gluon coherence is taken into account in the parton
shower via ordering of the opening angles of successive
branchings [\ref{marweb}-\ref{book}].

An alternative approach to cluster formation and decay is to use a
higher value of the cutoff $t_0$ and an anisotropic, multihadron decay
scheme for the resulting heavy clusters [\ref{Got}].  Clearly, this
approach lies somewhere between the low-mass cluster and string models.
In practice, even with a low value of $t_0$ one needs to invoke some such
decay scheme for the small fraction of clusters that have masses of more
that an few GeV/c$^2$, for which isotropic two-body decay is an
implausible hypothesis.

\subsection{QCD event generators}
There are three classes of programs for generating full events from
parton showers, using each of the three above-mentioned hadronization
models.

Because of the difficulties discussed in connection with the independent
fragmentation model, one would expect it
to work best for the higher-momentum hadrons in final states consisting of
a few well-separated jets.  Probably on account of its simplicity, it was
initially the model used most widely in conjunction with initial- and
final-state parton showering for the simulation of hard hadron-hadron
collisions, in the programs {\small ISAJET} [\ref{PP}],
{\small COJETS} [\ref{Odo}] and {\small FIELDAJET} [\ref{Fie}].

The string hadronization model outlined above, with many further
refinements, is the basis of the \JS\ simulation program
[\ref{JS}], which also includes final-state parton showering
with optional angular ordering.  This program gives a very good
detailed description of hadronic final states in \ee\ annihilation
up to the highest energies studied so far [\ref{LEPMC}].

The \JS\ hadronization scheme is also used, in combination with
initial- and final-state parton cascades, in the other very successful
Lund simulation programs {\small PYTHIA} [\ref{Pyth}] for hadron-hadron
and {\small LEPTO} [\ref{Lept}] for lepton-hadron collisions. The alternative
formulation of parton showers in terms of colour dipole splitting
mentioned above is implemented in the program {\small ARIADNE} [\ref{Ariadne}],
which also uses \JS\ for hadronization.

The program \HW\ [\ref{HW}] uses a low-mass cluster hadronization
model [\ref{clus}] in conjunction with initial- and final-state parton
cascades to simulate a wide variety of hard scattering processes.
The showering
algorithm includes angular ordering and azimuthal correlations due to
coherence and gluon polarization. This approach gives a good account
of diverse data with relatively few adjustable parameters [\ref{LEPMC}].

More detailed descriptions and comparisons of event generators for
\ee\ physics may be found in Ref.~[\ref{lepevgen}].

\section{Event shapes}
\label{sec_evt}
A popular way to study the jet-like characteristics of hadronic
final states in \ee\ annihilation is to use {\em event shape} variables.
The procedure is to define a quantity $X$ which measures some particular
aspect of the shape of the hadronic final states, for example whether
the distribution of hadrons is  pencil-like, planar, spherical etc.
The differential cross section $ d \sigma / d X$ can be measured and
compared with the theoretical prediction. For the latter to be
calculable in perturbation theory, the variable should be
{\em infrared safe}, i.e.\ insensitive to the emission of soft or
collinear particles.  This is because the QCD matrix elements have
singularities whenever a soft gluon or collinear pair of massless
partons is emitted.  In particular, if $\bm{p}_i $ is  any 3-momentum
occurring in its definition, $X$ must be invariant under the branching
\begin{equation}
\bm{p}_i \to \bm{p}_j + \bm{p}_k
\end{equation}
whenever $\bm{p}_j$ and $\bm{p}_k$ are parallel or one of them goes to
zero. Quantities made out of linear sums of momenta meet this
requirement. The most widely-used example is the
{\em thrust} [\ref{defthrust}]
\def\vpi{\bm{p}_i}
\def\vpj{\bm{p}_j}
\def\vn{\bm{n}}
\beq\label{thrus}
T = \max\, {  \sum_i \vert \vpi\cdot\vn \vert \over
                         \sum_i \vert \vpi \vert   }\;,
\eeq
where $\vpi$ are the final-state hadron (or parton) momenta
and $\vn$ is an arbitrary unit vector. If the $\vpi$ form an almost
collinear pair of jets, then after the specified maximization or
minimization $\vn$ will lie along the jets, defining the {\em thrust
axis} of the event. As the jets become more pencil-like, the
thrust approaches unity.

A event shape variable that does not require finding an event axis
is the {\em C-parameter} [\ref{ERT}]
\beq\label{cpar}
C = {3\over 2}\; \frac{  \sum_{i,j}\vert\vpi\vert\,\vert\vpj\vert\,
 \sin^2\theta_{ij} } {  (  \sum_i \vert \vpi \vert )^2  }\;.
\eeq
In the case of a pencil-like two-jet event, the $C$-parameter is
close to zero; the normalization is such that the maximum value
of $C$ is unity.

As mentioned in Sect.~\ref{sec_mod}.4, the QCD Monte Carlo event
generator programs \HW\ and \JS\ are quite successful in
describing the properties of \ee\ final states, including the
distributions of event shape variables [\ref{LEPMC}]. From these
successes, we can hope that the models built into those
programs provide some guidance on the broad features of
the hadronization process. What the programs suggest is that
corrections to event shapes are still substantial
at energy scales $Q\sim\sub{M}{Z}$, typically around 10\% of the
leading-order QCD predictions. This is
comparable with the next-to-leading $\ho 2$
terms. Furthermore hadronization effects fall off
rather slowly with increasing energy, apparently
like $1/Q$. Thus it is imperative to understand
hadronization better in order to reap the benefit of
any future $\ho 3$ calculations of event shapes.

\begin{figure}
  \centerline{
    \setlength{\unitlength}{1cm}
    \begin{picture}(0,8.5)
       \put(0,0){\includegraphics{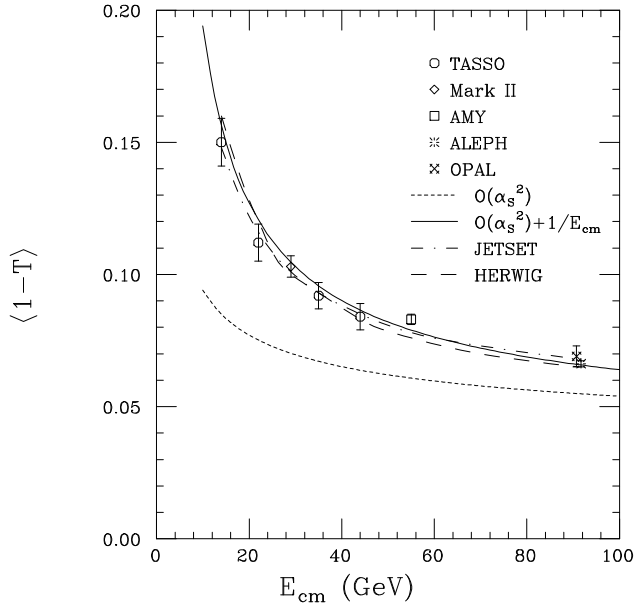}}
    \end{picture}}
  \caption[DATA]{
Mean value of 1 -- thrust in \ee\ annihilation.
 }\label{meant}
\end{figure}

We saw in Sect.~\ref{sec_mod} that $1/Q$ corrections are
in fact generated by the simplest `tube' model of hadronization.
Consider for example the thrust distribution. The thrust
of a two-jet event is precisely the jet momentum divided
by the jet energy, and therefore in the tube model
we expect a hadronization correction of $-2\lambda/Q$.
Thus for $\lambda\sim 500$ MeV the hadronization correction
to the thrust is expected to be
\beq\label{thad}
\sub{\VEV{\delta T}}{had}\sim -\frac{1\;\mbox{GeV}}{Q}\;.
\eeq
The purely perturbative
prediction for the mean thrust is [\ref{lepqcd}]
\beq
\VEV{1-T} = 0.335\as +1.02\as^2 + {\cal O}(\as^3)\; ,
\eeq
which implies that $\VEV{1-T}\sim 0.055$ at $Q=M_Z$, assuming
$\asZ\sim 0.12$. In fact, the value measured at LEP is
$\VEV{1-T} = 0.068\pm 0.003$, consistent with an additional
non-perturbative contribution of (1 GeV)$/Q$, which also agrees with
the energy dependence of $\VEV{1-T}$ down to about $Q=15$ GeV
(Fig.~\ref{meant}).

For the mean value of the $C$-parameter (\ref{cpar}), the tube model
predicts a hadronization correction of $3\pi\lambda/Q$, giving
\beq\label{chad}
\sub{\VEV{\delta C}}{had}\sim \frac{5\;\mbox{GeV}}{Q}\;.
\eeq
The perturbative prediction is
\beq
\VEV{C} = 1.374\as +4.00\as^2 + {\cal O}(\as^3)\; ,
\eeq
which implies $\VEV{C}\sim 0.222$ at $Q=M_Z$, again assuming
$\asZ\sim 0.12$. The value measured at LEP is
$\VEV{C} = 0.260\pm 0.006$, which is consistent
with the expected additional non-perturbative contribution,
although in this case there are no published lower-energy
measurements with which to check the $Q$-dependence.

\begin{figure}
  \centerline{
    \setlength{\unitlength}{1cm}
    \begin{picture}(0,11)
\put(0,0){\includegraphics{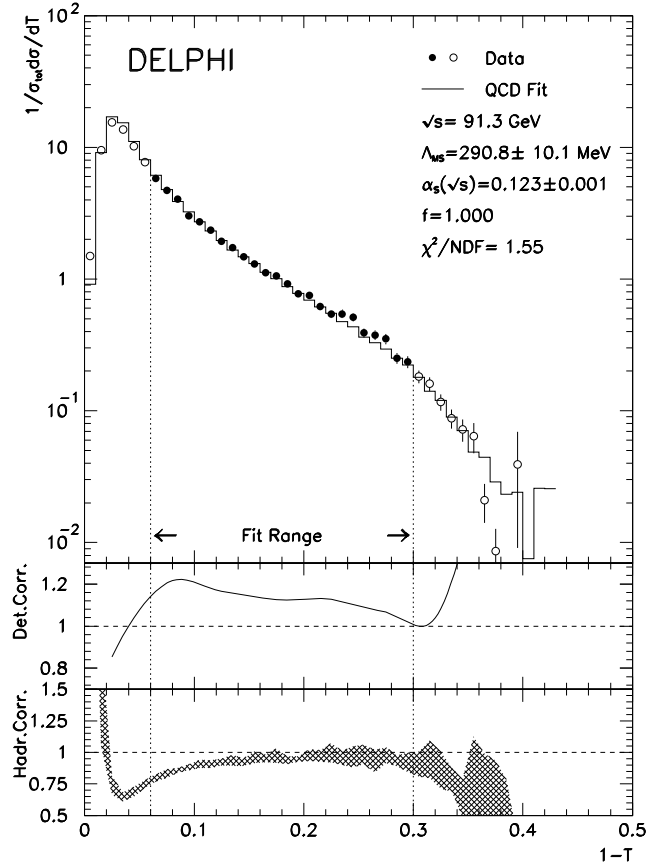}}
    \end{picture}}
  \caption[DATA]{Thrust distribution in \Z0\ $\to$ hadrons, with
detector and hadronization corrections.}
\label{fig_resum3}
\end{figure}

Turning to the differential distributions of event shape variables,
hadronization seems mainly to cause a smearing by an amount
proportional to $1/Q$. The effect is therefore most pronounced
in the neighbourhood of the peak of the distribution, as shown
for the thrust distribution in Fig.~\ref{fig_resum3}. The
shaded band shows the variation in the hadronization corrections
deduced from \JS\ and \HW, where the correction is defined
as the ratio of the parton- and hadron-level Monte Carlo predictions.
The predominant effect of hadronization is to smear out the
parton-level peak at low values of $1-T$, enhancing the hadron-level
distribution at intermediate values of the thrust.

\section{Jet fragmentation}
\label{sec_frag}
We have illustrated in Sect.~\ref{sec_evt} the use of perturbation
theory to calculate `infrared safe' quantities such
as \ee\ event shapes.  There are in addition
predictions that can be made concerning some quantities that
are `infrared sensitive', i.e.\ that have infrared and
collinear singularities in perturbation theory.  Such
quantities can still be handled provided the singularities
can be collected into an overall factor which describes the
sensitivity of the quantity to long-distance physics. The
divergence of this factor corresponds to the fact that
long-distance phenomena are not reliably predicted by
perturbation theory. Therefore the divergent factor
must be replaced by a finite factor determined either
by experiment or according to some non-perturbative
model. Once this is done, perturbation theory can be
used to predict the scale dependence of the quantity.

The best known example of a perturbative prediction concerning
a factorizable infrared-sensitive quantity is the phenomenon
of scaling violation in hadron structure functions. There one
studies the parton distributions inside a hadron,
probed by deep inelastic lepton scattering. Here we
consider the related phenomenon for the fragmentation of a
jet, produced for example in \ee\ annihilation, into hadrons
of a given type $h$.

\subsection{Fragmentation functions}
The total fragmentation function for hadrons of
type $h$ in \ee\ annihilation at c.m.\ energy $\rs$ as
is defined as
\beq\label{Fxs}
F^h(x,s) = \frac{1}{\st}\frac{d\sigma}{dx}
(e^+e^-\to h X)
\eeq
where $x=2E_h/\rs\le 1$ is the scaled hadron
energy.\footnote{In practice, the approximation
$x=x_p=2p_h/\rs$ is often used.} These
functions are predicted by the QCD event generators
discussed in Sect.~\ref{sec_mod}.4, and experimental
data for the various identifiable hadron species provide
strong constraints on the hadronization models used in
the programs.  Generally speaking, the most developed
event generators, \JS\ and \HW, give fairly good
agreement with experiment, although there are problems
with the yields of heavy strange particles [\ref{LEPMC}].

The fragmentation function (\ref{Fxs}) can be represented as
a sum of contributions from the different primary partons
$i=u,d,\ldots ,g$:
\beq\label{Fsum}
F^h(x,s) = \sum_i\int \frac{dz}{z} C_i(s;z,\as) D^h_i(x/z,s)\;.
\eeq
In lowest order the coefficient function $C_g$ for gluons is zero,
while for quarks $C_i=g_i(s)\delta(1-z)$ where $g_i(s)$ is the
appropriate electroweak coupling.

We cannot compute the parton $\to$ hadron fragmentation functions
$D^h_i$ from perturbation theory, since the production of hadrons is
not a perturbative process. We might consider trying to compute
functions $D^j_i$ that would describe the fragmentation of
partons of type $i$ into partons of type $j$. However, they
would be infinite, since the probability of emitting a
collinear gluon or light quark is divergent. Nevertheless,
it can be shown [\ref{factn}] that all such divergences are
factorizable in the sense that one can write
\beq\label{factFF}
D^j_i(x,t) = \sum_k \int_x^1
dz\,K^k_i(z,t,t_0)\,D^j_k(x/z,t_0)
\eeq
where the kernel function $K^k_i$ is perturbatively
calculable and $K^k_i(z,t,t)=\delta_{ik}\,\delta(1-z)$.

In the real world, hadrons are formed and fragmentation
functions are not divergent. As long as the scales $t$ and
$t_0$ are large, we would not expect the form of
Eq.~(\ref{factFF}) to be affected by such long-distance
phenomena.  Therefore we replace $j$ by $h$ and apply the
equation to hadronic fragmentation functions. After measuring
these functions at some scale $t_0$, we can use the equation
to predict their form at any other scale $t$, as long as both
$t$ and $t_0$ are large enough for perturbation theory to be
applicable in the calculation of the kernel $K^k_i$.

\subsection{Scaling violation in jet fragmentation}
Consider the change in a fragmentation function $D_i(x,t)$
when the hard process scale is increased from $t$ to $t+dt$.
Such a change can only occur via the splitting of a parton of
type $i$ in this scale interval.  Hence the fragmentation
functions $D_i$ satisfy evolution equations of the Altarelli-Parisi
type, first introduced for deep inelastic scattering [\ref{AlP}]:
\beq\label{APeq}
\tdt D_i(x,t) = \sum_j\int \frac{dz}{z} \frac{\as}{2\pi}
P_{ji}(z,\as) D_j(x/z,t)\;.
\eeq
The $i\to j$ parton splitting function $P_{ji}$ has a perturbative
expansion of the form
\beq\label{Pijexp}
P_{ji}(z,\as) =P^{(0)}_{ji}(z)
+\frac{\as}{2\pi} P^{(1)}_{ji}(z) +\cdots
\eeq
where the lowest-order splitting function $P^{(0)}_{ji}(z)$
is the same in fragmentation and deep inelastic scattering
but the higher-order terms are different. The effect
of splitting is qualitatively the same in both cases: as the
scale increases, one observes a {\em scaling violation} in
which the $x$ distribution is shifted towards lower values.

The most common strategy for solving the evolution
equations (\ref{APeq}) is to take moments (Mellin transforms)
\index{Mellin transform} with respect to $x$:
\beq\label{Mellin}
\tilde D(N,t) = \int_0^1 dx\; x^{N-1}\; D(x,t)\; ,
\eeq
the inverse Mellin transform being
\beq\label{invMell}
D(x,t) = \frac{1}{2\pi i}\int_C dN\; x^{-N}\; \tilde D(N,t)\;,
\eeq
where the integration contour $C$ in the complex $N$ plane
is parallel to the imaginary axis and to the right of all
singularities of the integrand.  After Mellin transformation,
the convolution on the right-hand side of Eq.~(\ref{APeq})
becomes simply a product.

The moments $\tilde P_{ji}$ of the splitting functions
are called {\em anomalous dimensions}, usually denoted by
$\gamma_{ji}(N,\as)$. In lowest order they take the form
\beeq \label{z77}
\gamma^{(0)}_{qq}(N) &=&
 C_F \Bigg[ -\frac{1}{2}
+\frac{1}{N(N+1)} -2\sum^N_{k=2} \frac{1}{k} \Bigg] \nonumber \\
\gamma^{(0)}_{qg}(N) &=&
T_R \Bigg[ \frac{(2+N+N^2)}{N(N+1)(N+2)} \Bigg] \nonumber \\
\gamma^{(0)}_{gq}(N) &=&
C_F \Bigg[ \frac{(2+N+N^2)}{N(N^2-1)} \Bigg] \\
\gamma^{(0)}_{gg}(N) &=&
2C_A \Bigg[ -\frac{1}{12} +\frac{1}{N(N-1)} +\frac{1}{(N+1)(N+2)}
- \sum_{k=2}^N \frac{1}{k} \Bigg]   -\frac{2}{3} \nlf T_R \;, \nonumber
\eeeq
where the QCD colour factors are $C_F=\rat 4 3$, $C_A=3$,
$T_R=\half$, and $\nlf$ is the number of quark flavours
with masses less than the relevant scale $\sqrt{t}$.

We can consider fragmentation function
combinations which are non-singlet in flavour space, such as
$D_V=D_{q_i} - D_{\bar q_i}$ or $D_{q_i} -D_{q_j}$. In these
combinations the mixing with the flavour singlet gluons drops
out and for a fixed value of $\as$ the solution is simply
\beq
\tilde D_V(N,t) = \tilde D_V(N,t_0)
\left(\frac{t}{t_0}\right)^{\gamma_{qq}(N,\as)}  \;.
\eeq
For a running coupling $\as(t)$, the scaling violation is no
longer power-behaved in $t$. Inserting the lowest-order form
for the running coupling,
\beq\label{aslo}
\as(t) = \frac{1}{b\ln(t/\Lambda^2)}
\eeq
where $b=(11C_A-2\nlf)/12\pi$, we find the solution
\beq \label{scaviol}
\tilde D_V(N,t) = \tilde D_V(N,t_0) \left(\frac{\as
(t_0)}{\as (t)}\right)^{d_{qq}(N)} 	, \;\; d_{qq}(N)=
\frac{\gamma^{(0)}_{qq} (N)}{2\pi b}\; ,
\eeq
which varies like a power of $\ln t$.

For the singlet quark fragmentation function
\beq
D_S=\sum_i (D_{q_i} + D_{\bar q_i})\; ,
\eeq
we have mixing with the fragmentation of the gluon and the
evolution equation becomes a matrix relation of the form
\beq\label{singev}
\tdt \left(\begin{array}{c} \tilde D_S \\ \tilde D_g \end{array}\right)
= \left(\begin{array}{cc} \gamma_{qq} & 2\nlf\gamma_{gq} \\
\gamma_{qg} & \gamma_{gg} \end{array}\right)
\left(\begin{array}{c} \tilde D_S \\ \tilde D_g \end{array}\right)\; .
\eeq
The anomalous dimension matrix in this equation has two
real eigenvalues $\gamma_\pm$ given by
\beq
\gamma_{\pm}=\half[\gamma_{gg}+\gamma_{qq}\pm
\sqrt{(\gamma_{gg}-\gamma_{qq})^2+8\nlf\gamma_{gq}\gamma_{qg}}]\;.
\eeq
Expressing $D_S$ and $D_g$ as linear combinations of the
corresponding eigenvectors $D_+$ and $D_-$, we find that they
evolve as superpositions of terms of the form (\ref{scaviol})
with $\gamma_+$ and $\gamma_-$ in the place of $\gamma_{qq}$.

At small $x$, corresponding to $N\to 1$, the
$g\to g$ anomalous dimension becomes dominant and we find
$\gamma_+\to \gamma_{gg}\to\infty$, $\gamma_-\to\gamma_{qq}\to 0$.
This region requires special treatment, as we discuss in the
following two sections.

There are several complications in the experimental study of
scaling violation in jet fragmentation [\ref{NW}].
First, the energy dependence of the electroweak couplings
$g_i(s)$ that enter into Eq.~(\ref{Fsum})
is especially strong in the energy region presently under study
($\rs =20-90$ GeV).  In particular, the $b$-quark contribution
more than doubles in this range.
The fragmentation of the $b$ quark into
charged hadrons, including the decay products of the
$b$-flavoured hadron, is expected to be substantially softer
than that of the other quarks, so its increased contribution
can give rise to a `fake' scaling violation that has nothing to
do with QCD. A smaller, partially compensating effect is
expected in charm fragmentation. These effects can be
eliminated by extracting the $b$ and $c$ fragmentation
functions at $\rs = \sub{M}{Z}$ from tagged heavy quark
events, and evolving them separately to other energies.

Secondly, one requires the gluon fragmentation function $D_g(x,s)$
in addition to those of the quarks.  Although the gluon does not
couple directly to the electroweak current, it contributes
in higher order, and mixes with the quarks through evolution.
Its fragmentation can be studied in tagged heavy-quark
three-jet ($Q\bar Q g$) events, or via the longitudinal
fragmentation function (see Sect.~\ref{sec_frag}.5).

A final complication is that power corrections to
fragmentation functions, of the form $f(x)/Q^p$,
are not well understood.  As in the case of event shapes,
Monte Carlo studies [\ref{NW}] suggest that hadronization
can lead to $1/Q$ corrections. Therefore,
possible contributions of this form should be
included in the parametrization
when fitting the scaling violation.

\begin{figure}
  \centerline{
    \setlength{\unitlength}{1cm}
    \begin{picture}(0,11)
       \put(0,0){\includegraphics{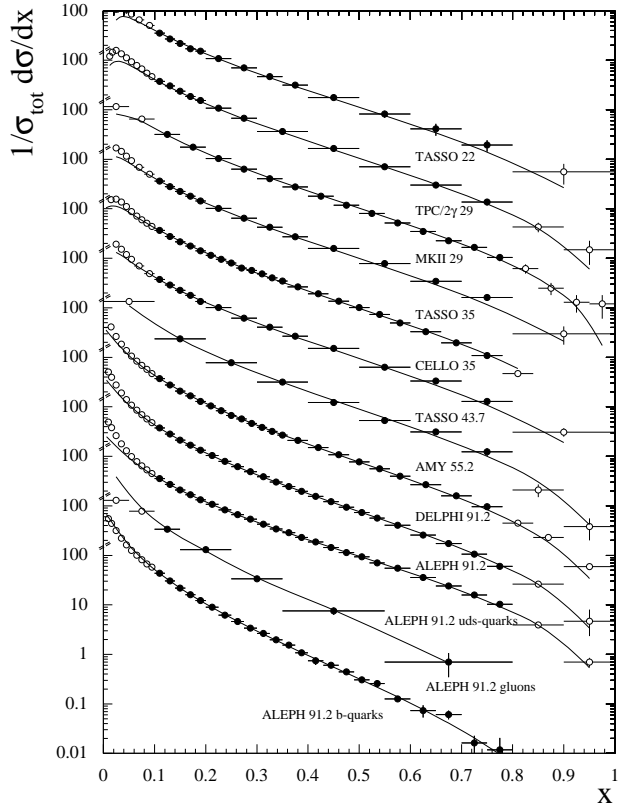}}
    \end{picture}}
  \caption[DATA]{
Scaling violation in \ee\ fragmentation functions.
 }\label{aleph_scav}
\end{figure}

Preliminary results of an analysis of scaling
violation in the charged hadron spectrum by the ALEPH
collaboration [\ref{scaALEPH}], based on a comparison
of LEP data with those from lower-energy experiments,
are shown in Fig.~\ref{aleph_scav}. Also included
are their separate fragmentation functions for
light ($u,d,s$) quarks, $b$ quarks and gluons, showing
that the latter two functions are significantly softer
(quite similar to each other, within the errors)
at $\rs = \sub{M}{Z}$.  An overall fit in the
range $22\leq\rs\leq 91.2$ GeV, $0.1<x<0.8$,
incorporating full next-to-leading-order evolution
and a simple parametrization of $1/Q$ power corrections,
gives a good description of the data in the fitted
region, as shown by the curves.\footnote{The fitted
curves do not extrapolate well into the small-$x$ region,
but a next-to-leading order treatment would not be
expected to be reliable there: see Sect.~\ref{sec_frag}.4.}
The fitted power corrections are small and the value obtained
for $\asZ$ ($0.127\pm 0.011$) is not highly sensitive
to the form assumed for them. Indeed, a $1/Q^2$ dependence
is not ruled out, and it may be that in this case
the Monte Carlo models are misleading and $1/Q$
corrections are in fact absent.
An operator-product approach to fragmentation,
developed in Ref.~[\ref{BalitskyBraun}], does
indeed suggest that 1/Q corrections are absent.

\subsection{Average multiplicity}
The average number of hadrons of
type $h$ in a jet initiated by a parton $i$ at scale $t$,
$\cN^h_i(t)$, is just the integral of the fragmentation
function, which is the $N=1$ moment in the notation of
Eq.~(\ref{Mellin}):
\beq
\cN^h_i(t) = \int_0^1 dx\, D^h_i(x,t) = \tilde D^h_i(1,t)\; .
\eeq
If we try to compute the $t$ dependence of this quantity, we
immediately encounter the problem that the anomalous
dimensions $\gamma_{gq}$ and $\gamma_{gg}$ given in Eq.~(\ref{z77})
have poles at $N=1$. The reason is that for $N \le 1$ the moment
integrals are dominated by the region of small $z$, where $P_{gi}(z)$
has a divergence associated with soft gluon emission.

In fact, however, we can still solve the evolution equation
for the average multiplicity provided we take into account
the suppression of soft gluon emission due to QCD coherence
[\ref{DKT},\ref{book}]. The leading effect of coherence is obtained
simply by changing the evolution variable from virtual mass-squared
to the quantity $t=E^2[1-\cos\theta]$, where $\theta$ is the
opening angle, and imposing {\em angular ordering},
$\theta' <\theta$ [\ref{DKT},\ref{angord}].  Thus in terms
of the newly-defined evolution variable the ordering condition
is $t'<z^2 t$ and Eq.~(\ref{APeq}) becomes
\beq\label{APtlt}
\tdt D_i(x,t) = \sum_j\int_x^1\frac{dz}{z} \frac{\as}{2\pi}
P_{ji}(z,\as) D_j(x/z,z^2 t)\;.
\eeq
Notice that this differs from the conventional Altarelli-Parisi
equation only in the $z$-dependent change
of scale on the right-hand side.  This change is not
important for most values of $x$ but we shall see that it is
crucial at small $x$.

For simplicity, consider first the solution of
Eq.~(\ref{APtlt}) taking $\as$ fixed and neglecting the sum
over different branchings.  Then taking moments as usual we
have
\beq\label{APtltN}
\tdt\tilde D(N,t) =
\frac{\as}{2\pi}\int_x^1 dz\,z^{N-1} P(z)
\tilde D(N,z^2t)\;.
\eeq
Now if we try a solution of the form
\beq\label{FNtlt}
\tilde D(N,t) \propto t^{\gamma(N,\as)}
\eeq
we find that the anomalous dimension $\gamma(N,\as)$ must satisfy
the implicit equation
\beq\label{gamN}
\gamma(N,\as) = \frac{\as}{2\pi}\int_0^1 dz\,z^{N-1+2\gamma(N,\as)}
P(z)\; .
\eeq

When $N-1$ is not small, we may neglect the $2\gamma(N,\as)$ in
the exponent of Eq.~(\ref{gamN}) and then we obtain the usual
explicit formula for the anomalous dimension.  For $N\simeq
1$, the region we are interested in, the $z\to 0$ behaviour
of $P(z)$ dominates and we may as a first approximation keep
only those terms that are singular at $z=0$. The most
important such term appears in the gluon-gluon splitting
function: $P_{gg}(z) \to 2C_A/z$ as $z\to 0$. Then
Eq.~(\ref{gamN}) implies that near $N=1$
\beq
\gamma_{gg}(N,\as) = \frac{C_A\as}{\pi}
\frac{1}{N-1+2\gamma_{gg}(N,\as)}\;,
\eeq
and hence
\beeq\label{gamggN}
& &\gamma_{gg}(N,\as) \;=\;\frac{1}{4}\left[\sqrt{(N-1)^2 +
\frac{8C_A\as}{\pi}} - (N-1)\right]\nonumber \\
& & = \sqrt{\frac{C_A\as}{2\pi}} -\frac{1}{4}(N-1)
+\frac{1}{32}\sqrt{\frac{2\pi}{C_A\as}}(N-1)^2 + \cdots\;.
\eeeq

Thus for $N\to 1$ the gluon-gluon anomalous dimension
behaves like the square root of $\as$.  How can this
behaviour emerge from perturbation theory, which deals only
in integer powers of $\as$?  The answer is that at any
fixed $N\neq 1$ we can expand Eq.~(\ref{gamggN}) in a different
way for sufficiently small $\as$:
\beq\label{gamexp}
\gamma_{gg}(N,\as) = \frac{C_A\as}{\pi(N-1)}-
2\left(\frac{C_A\as}{\pi}\right)^2 \frac{1}{(N-1)^3}
+\cdots \;.
\eeq
This series displays the terms that are most singular as
$N\to 1$ in each order.  These terms have been {\em resummed}
in the expression (\ref{gamggN}), allowing the perturbation
series to be analytically continued outside its circle of
convergence $|\as|<(\pi/8C_A)|N-1|^2$.  By definition, the
behaviour outside this circle (in particular, for the
average multiplicity, at $N=1$) cannot be represented
by the series any more, even though it is fully implied by it.

At sufficiently small $x$, the $N\to 1$ singularity of the
gluon-gluon anomalous dimension dominates in all
fragmentation functions. Thus we obtain the behaviour (\ref{FNtlt})
with $\gamma =\gamma_{gg}$ for the total fragmentation
function defined in Eq.~(\ref{Fsum}). To predict this behaviour
quantitatively we need to take account of the running of $\as$,
which can be done in the same way as for the other moments. Writing
Eq.~(\ref{FNtlt}) in the form
\beq\label{tilFsimD}
\tilde F(N,t) \sim\tilde D(N,t)
\propto\exp\left[\int^{t} \gamma_{gg}(N,\as) \frac{dt'}{t'}\right] \;,
\eeq
we have to replace $\gamma_{gg}(N,\as)$ in the integrand
by $\gamma_{gg}(N,\as(t'))$.  We then write
\beq
\int^{t} \gamma_{gg}(N,\as(t')) \frac{dt'}{t'}
= \int^{\as(t)}\frac{\gamma_{gg}(N,\as)}{\beta(\as)}
\,d\as\;,
\eeq
where $\beta(\as) = -b\as^2+\cdots$, and find
\beeq\label{FNasy}
\tilde F(N,t)&\sim & \exp\Biggl[
 \frac{1}{b}\sqrt{\frac{2C_A}{\pi\as}} -\frac{1}{4b\as}(N-1)
\nonumber \\
&+& \frac{1}{48b} \sqrt{\frac{2\pi}{C_A\as^3}}(N-1)^2
+\cdots\Biggr]_{\as=\as(t)}\;.
\eeeq

In \ee\ annihilation the scale $t$ (the upper limit on
$E^2[1-\cos\theta]$ for any branching) is of the order of the
centre-of-mass energy-squared $s$, and so the average
multiplicity of any hadronic species has the asymptotic
behaviour
\beq\label{avmult}
\VEV{n(s)} = \tilde F(1,s) \sim \exp
\frac{1}{b}\sqrt{\frac{2C_A}{\pi\as(s)}} \sim
\exp\sqrt{\frac{2C_A}{\pi b}
\ln\left(\frac{s}{\Lambda^2}\right)}\;.
\eeq

\begin{figure}
\vspace{10cm}
\includegraphics{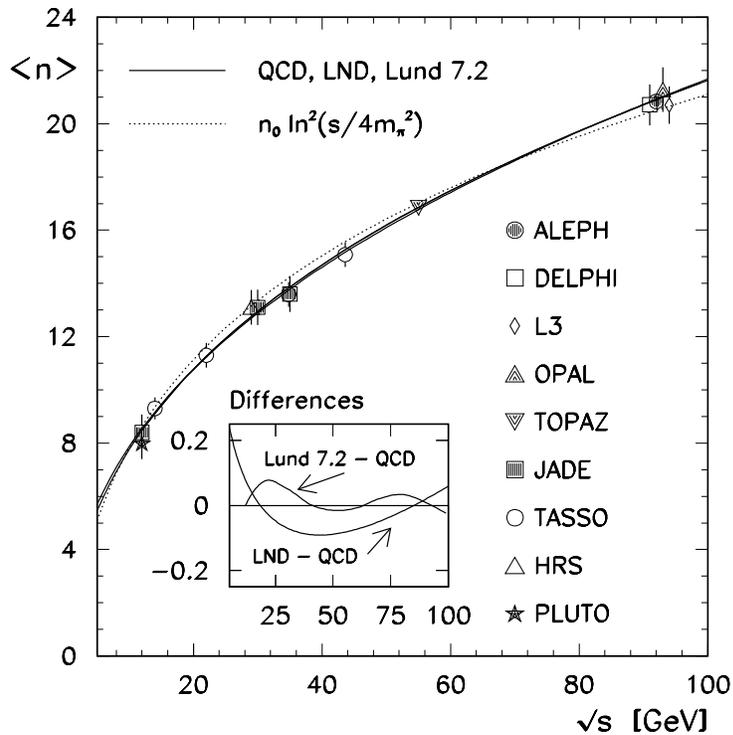}
\caption{Average multiplicity of charged hadrons in \ee\ annihilation.}
\label{fig_avmult}
\end{figure}

If the next-to-leading singularities of the anomalous dimensions,
i.e.\ the terms with one less power of $1/(N-1)$ in
Eq.~(\ref{gamexp}), are also resummed, the expression (\ref{avmult})
is multiplied by a power of $\as(s)$ [\ref{Mueller}]. The resulting
prediction, shown by the solid curve in Fig.~\ref{fig_avmult}, is
in very good agreement with experiment [\ref{mulex}].

\subsection{\boldmath Small-$x$ fragmentation}
\begin{figure}
\vspace{10cm}
\includegraphics{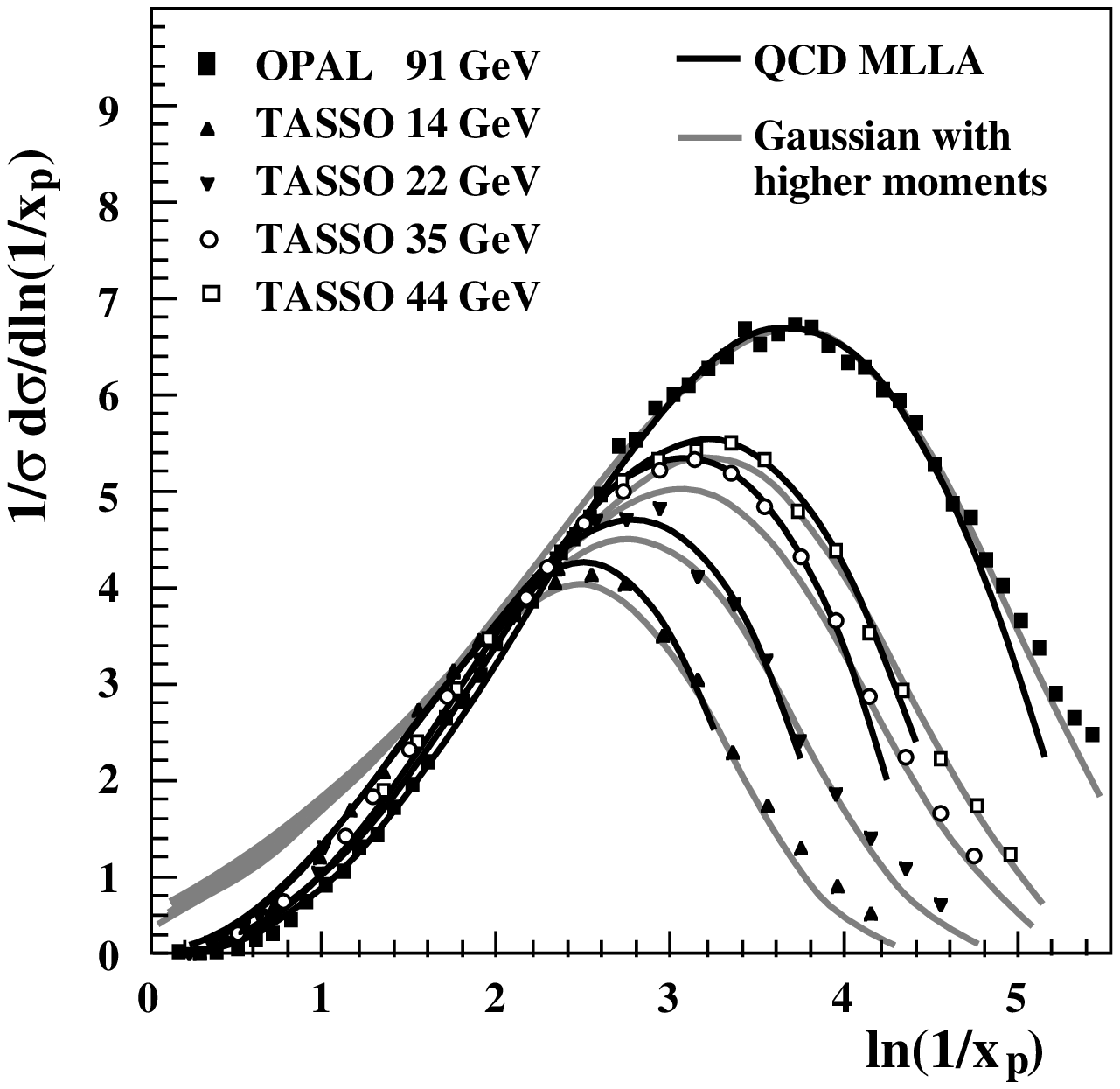}
\caption{Distribution of $\xi=\ln(1/x)$ in \ee\ annihilation.}
\label{fig_hump}
\end{figure}

The behaviour of $\tilde F(N,s)$ away from $N=1$
determines the form of small-$x$
fragmentation functions via the inverse Mellin transformation
(\ref{invMell}). Keeping the first three terms in the Taylor
expansion of the exponent, as displayed in
Eq.~(\ref{FNasy}), gives a simple Gaussian function of $N$
which transforms into a Gaussian in the variable
$\xi\equiv\ln(1/x)$:
\beq\label{gaussxF}
xF(x,s)\propto \exp\left[-\frac{1}{2\sigma^2}(\xi-
\xi_p)^2\right]\;,
\eeq
where the peak position is
\beq\label{xipeak}
\xi_p = \frac{1}{4b\as(s)}\sim \frac{1}{4}\ln s
\eeq
and the width of the distribution of $\xi$ is
\beq\label{xiwidth}
\sigma = \left(\frac{1}{24b}\sqrt{\frac{2\pi}{C_A\as^3(s)}}
\right)^{\frac{1}{2}} \propto (\ln s)^{\frac{3}{4}}\;.
\eeq
Again, one can compute next-to-leading corrections to these
predictions [\ref{FongWeb}], and the results agree very well with
the form and energy dependence of fragmentation functions at small
$x$ as measured in \ee\ annihilation (Fig.~\ref{fig_hump}) [\ref{humpex}].
This provides support for the notion of local parton-hadron duality
mentioned in Sect.~\ref{sec_mod}. We assumed in Eq.~(\ref{tilFsimD})
that the $N$-dependence of the Mellin-transformed fragmentation
function is dominated by the perturbative part involving the
anomalous dimension, and that the non-perturbative factor
is relatively smooth. Smoothness in $N$ transforms into
locality in $x$. Although the violation of scaling ensures
that this is valid asymptotically, it seems to be a good
approximation already at energies as low as 14 GeV.

\begin{figure}
\vspace{8.0cm}
\includegraphics{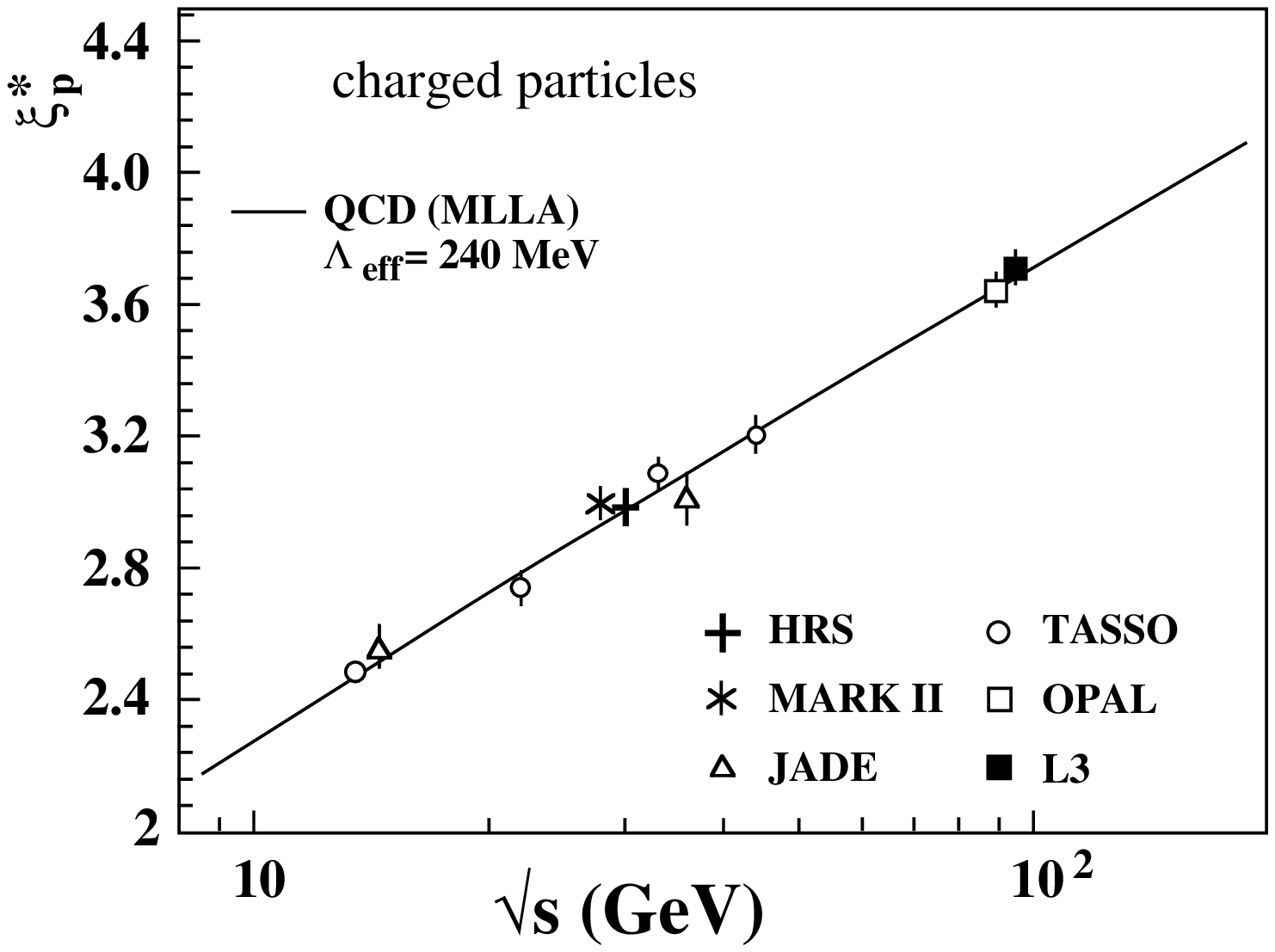}
\caption{Position of the peak in the $\xi$-distribution.}
\label{fig_peak}
\end{figure}

The energy-dependence (\ref{xipeak}) of the peak in the
$\xi$-distribution is also a striking illustration of soft gluon
coherence, which is the origin of the suppression of hadron production
at small $x$. Of course, a decrease at very small $x$ is
expected on purely kinematical grounds, but this would occur
at particle energies proportional to their masses, i.e.\ at
$x\propto m/\sqrt{s}$ and hence $\xi\sim\frac{1}{2}\ln s$.
Thus if the suppression were purely kinematic the peak
position $\xi_p$ would vary twice as rapidly with energy,
which is ruled out by the data (Fig.~\ref{fig_peak}).

\subsection{Longitudinal fragmentation}
Recently, the ALEPH [\ref{scaALEPH}] and
OPAL [\ref{longOPAL}] collaborations at LEP have
presented preliminary results of the first analyses
of the longitudinal fragmentation function in
\ee\ annihilation.
This is defined in terms of the joint distribution in the
energy fraction $x$ and the angle $\theta$ between the observed
hadron and the incoming electron beam [\ref{NW}]:
\beq\label{sigTLA}
\frac{1}{\st}\frac{d^2\sigma}{dx\,d\cos\theta}
= \frac 3 8 (1+\cos^2\theta)\,\sub F T(x)
+ \frac 3 4    \sin^2\theta \,\sub F L(x)
+ \frac 3 4    \cos  \theta \,\sub F A(x)\; ,
\eeq
where $\sub F T$, $\sub F L$ and $\sub F A$ are respectively the
transverse, longitudinal and asymmetric fragmentation
functions.\footnote{All these functions also depend
on the c.m.\ energy $\rs$, which we take to be fixed here.}
Like the total fragmentation function,
$F=\sub F T +\sub F L$,
each of these functions can be represented as a convolution
of the parton fragmentation functions $D_i$ with appropriate
coefficient functions $C_i^{\mbox{\scriptsize T,L,A}}$ [\ref{coeffns}]
as in Eq.~(\ref{Fsum}). In fact the transverse and longitudinal
coefficient functions are related in such a way that
\beq\label{FL}
\sub F L(x) = \frac{\as}{2\pi}C_F\int\frac{dz}{z}
\Bigl[ \sub F T(z)
+ 4\left(\frac z x -1\right)D_g(z)\Bigr] + \ho 2\;.
\eeq
Thus the gluon fragmentation function $D_g$ can be extracted
to leading order from measurements of $\sub F T$ and $\sub F L$.

\begin{figure}
  \centerline{
    \setlength{\unitlength}{1cm}
    \begin{picture}(0,8)
       \put(0,0){\includegraphics{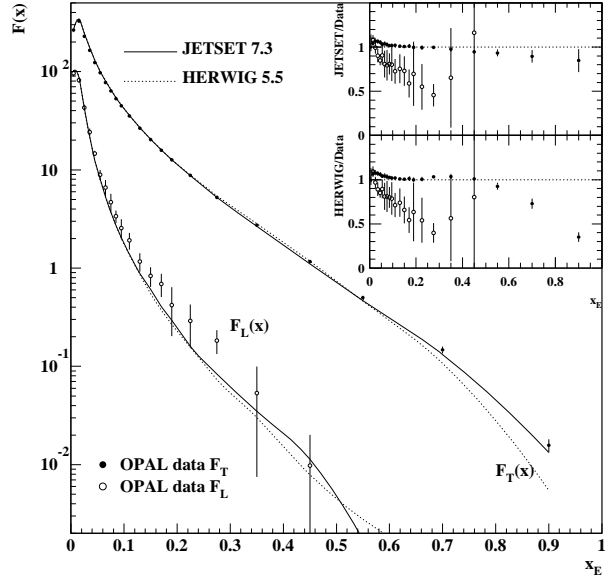}}
    \end{picture}}
  \caption[DATA]{
Transverse and longitudinal fragmentation functions in \Z0\ $\to$ hadrons.
 }\label{opal_ftl}
\end{figure}

\begin{figure}
  \centerline{
    \setlength{\unitlength}{1cm}
    \begin{picture}(0,8)
       \put(0,0){\includegraphics{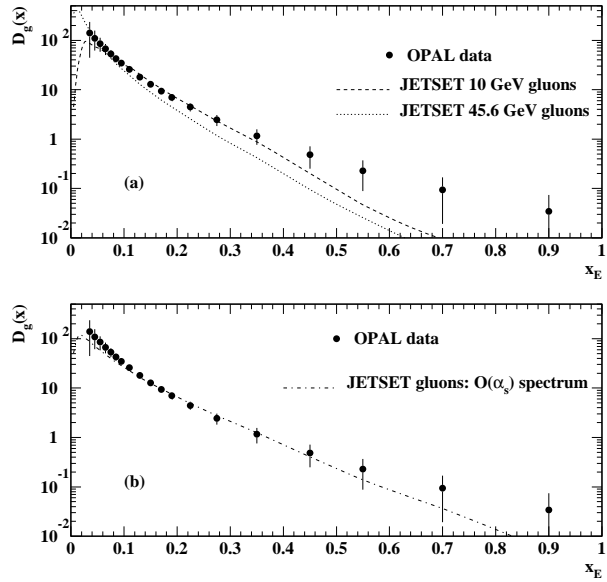}}
    \end{picture}}
  \caption[DATA]{
Gluon fragmentation function extracted from $\sub{F}{T,L}$.
 }\label{opal_dg}
\end{figure}

Fig.~\ref{opal_ftl} shows the preliminary OPAL data [\ref{longOPAL}]
on the transverse and longitudinal fragmentation functions
for charged particles. Note from Eq.~(\ref{FL}) that
$\sub F L$ is $\oas$ relative to $\sub F T$;
it also falls more steeply with $x$.  Therefore even with
LEP statistics the errors are large for $x>0.3$.
However, one still obtains useful information on the gluon
fragmentation function over the full range of $x$, as shown
in Fig.~\ref{opal_dg}. Because the relation (\ref{FL}) is
known only to leading order, there is an ambiguity about
the energy scale at which $D_g$ is measured. Comparisons
with \JS\ predictions at various scales are shown.

Similar, although systematically somewhat lower, preliminary
results on the longitudinal fragmentation function have been
obtained by the ALEPH collaboration [\ref{scaALEPH}], who include
this information in their fit to scaling violation as a further
constraint on the gluon fragmentation function shown in
Fig.~\ref{aleph_scav}.

Summed over all particle types, the total fragmentation function
satisfies the energy sum rule
\beq
\frac 1 2\int dx\,xF(x) = 1\;.
\eeq
Similarly the integrals
\beq
\frac 1 2\int dx\,x\sub F {T,L}(x) \equiv \frac{\sub\sigma{T,L}}{\st}
\eeq
tell us the transverse and longitudinal fractions of the total
cross section.  The perturbative prediction is
\beq\label{sigl}
\frac{\sub\sigma{L}}{\st} = \frac{\as}{\pi} + \ho 2\;,
\eeq
that is, the whole of the $\oas$ correction to $\st$ comes from
the longitudinal part. Surprisingly, the $\ho 2$ correction has
not yet been calculated.  Once this has been done, Eq.~(\ref{sigl})
will provide an interesting new way of measuring $\as$.

\begin{figure}
  \centerline{
    \setlength{\unitlength}{1cm}
    \begin{picture}(0,7.5)
       \put(0,0){\includegraphics{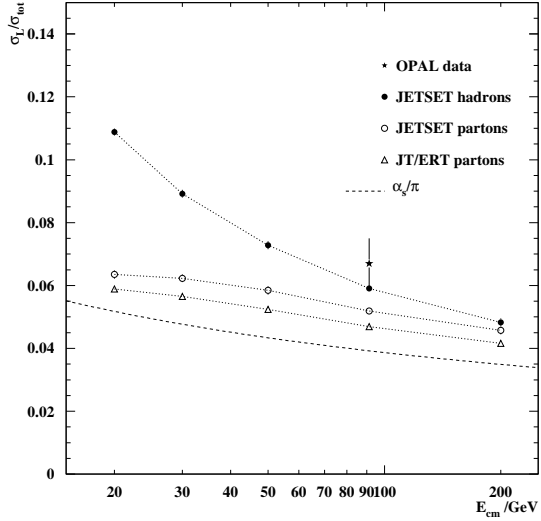}}
    \end{picture}}
  \caption[DATA]{
Longitudinal fraction of the \ee\ hadronic cross section.
 }\label{opal_sigl}
\end{figure}

The preliminary OPAL data point for $\sub{\sigma}{L}/\st$ is shown,
together with \JS\ Monte Carlo predictions, in Fig.~\ref{opal_sigl}.
It should be noted that neither of the \JS\ parton-level predictions
fully includes the $\ho 2$ contribution [\ref{torbjorn}]. We see,
however, that the data
lie well above the leading-order prediction (dashed), which
suggests that higher-order and/or non-perturbative corrections
are significant. An estimate of the latter is provided by the difference
between the \JS\ curves for hadrons and partons. This difference
shows a clear $1/Q$ behaviour, with a coefficient of about
1 GeV, illustrating once again the characteristic energy
dependence of hadronization effects.

In fact, in the simple `tube' model of hadronization discussed in
Sect.~\ref{sec_mod}, the correction to the longitudinal cross
is found to be
\beq\label{lhad}
\sub{\VEV{\frac{\delta\sub{\sigma}{L}}{\st}}}{had}
= \frac{\pi\lambda}{2Q}\sim \frac{0.8\;\mbox{GeV}}{Q}\;,
\eeq
which agrees well with the \JS\ prediction shown in
Fig.~\ref{opal_sigl}.  The correction arises from mixing
between the transverse and longitudinal angular dependences
in Eq.~(\ref{sigTLA}) due to hadronization.  The transverse cross
section receives an equal and opposite correction, and so there
is of course no $1/Q$ term in the total cross section.

\subsection{Jet fragmentation in deep inelastic scattering}
Jet physics in ep collisions promises to be a rich field of study
for the new HERA collider at DESY. One can for example compare the
properties of jets in \ee\ and ep collisions, which is a good
test of the factorizability of jet fragmentation functions
discussed above.

The appropriate comparison is between one hemisphere of an
\ee\ event, defined with respect to the thrust axis, and
the ``current jet hemisphere" of deep inelastic scattering
in the Breit frame of reference [\ref{Fey},\ref{DKT}]. The
Breit frame is the one in which the momentum transfer from
the lepton, $q^\mu$, has only a $z$-component: $q^\mu = (0,0,0,Q)$.
In the parton model, in this frame the struck quark enters
from the right with $z$-momentum $p_z = -\rat 1 2 Q$ and sees
the exchanged virtual boson as a ``brick wall", from which
it simply rebounds with $p_z = +\rat 1 2 Q$. Thus the
right-hand hemisphere of the final state should look just like
one hemisphere of an \ee\ annihilation event at $\sub{E}{cm}=Q$.
The left-hand (``beam jet") hemisphere, on the other hand,
is different from \ee\ because it contains the proton remnant,
moving with momentum $p_z = -(1-x)Q/2x$ in this frame.

\begin{figure}
  \centerline{
    \setlength{\unitlength}{1cm}
    \begin{picture}(0,8)
       \put(0,0){\includegraphics{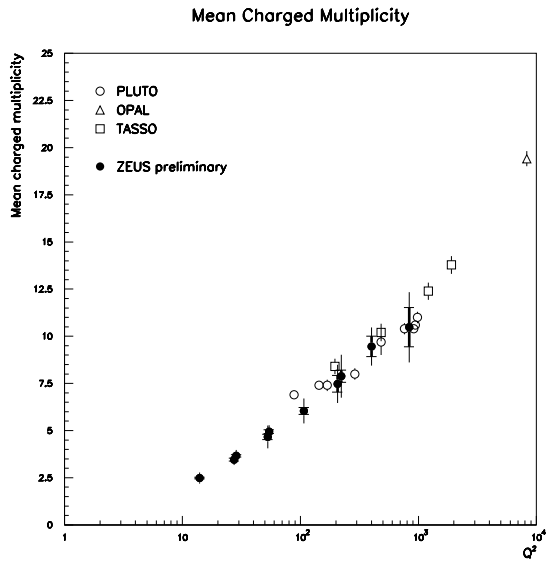}}
    \end{picture}}
  \caption[DATA]{
Mean charged multiplicity in \ee\ and ep current jet fragmentation.
 }\label{val_fig1}
\end{figure}

Fig.~\ref{val_fig1} shows the preliminary results of such a
comparison by the ZEUS collaboration [\ref{zeus}],
in this case comparing the charged multiplicity in the Breit
frame current hemisphere with that in \ee\ annihilation,
already shown in Fig.~\ref{fig_avmult}.
We see that the two sets of data do seem to follow the same curve.

\begin{figure}
  \centerline{
    \setlength{\unitlength}{1cm}
    \begin{picture}(0,8)
       \put(0,0){\includegraphics{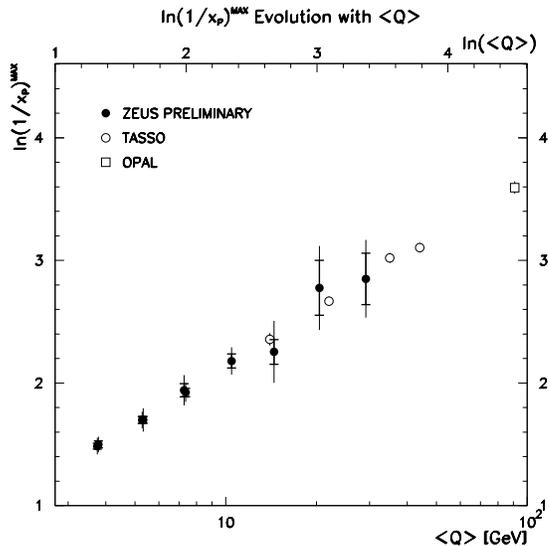}}
    \end{picture}}
  \caption[DATA]{
Position of the peak in $\ln(1/x_p)$ in \ee\ and ep current jet fragmentation.
 }\label{val_fig2}
\end{figure}

A more elaborate comparison is shown in Fig.~\ref{val_fig2}.
Here the position of the peak in the distribution of $\ln(1/x_p)$,
where $x_p=2p/Q$ in the Breit frame, is compared with that found
in \ee\ annihilation (Fig.~\ref{fig_peak}). We see again that the
preliminary HERA data follow and extrapolate the \ee\ curve,
in remarkable agreement with the theoretical predictions
derived in Sect.~\ref{sec_frag}.4.

\section{Power corrections from perturbation theory?}
\label{sec_pow}
We have seen that model studies suggest that for many infrared safe
quantities, such as \ee\ event shapes, hadronization corrections are
proportional to $1/Q$ where $Q$ is the centre-of-mass
energy.  This is in contrast to the \ee\ total cross section, which for
massless quarks has a leading power correction of order $1/Q^4$.  The
smallness of non-perturbative effects in the total cross section and
related quantities has in fact led to a preference for such quantities
as a means of determining $\as$, even though event shapes have a stronger
perturbative dependence on $\as$.

In the case of the total cross section, we also have some understanding
of the leading power correction [\ref{SVZ}]. It is believed to arise from the
vacuum expectation value of the gluon condensate, $\VEV{\as G^2}$, which is
the relevant gauge-invariant operator of lowest dimension, giving a
correction proportional to  $\VEV{\as G^2}/Q^4$.  From this viewpoint,
we do not even know why the corrections to event shapes should be of order
$1/Q$: there are no local operators of dimension one to which they could
be related.

A possibly fruitful way of discussing power corrections is in terms
of renormalons [\ref{ConSt}-\ref{brw}]. These are singularities of the
Borel transform of the all-orders perturbative expression for a quantity,
generated by factorial growth of the perturbation series at high
orders. The idea of Borel transformation is that for any perturbation
series
\beq
P(\as) = \sum_{n=0}^\infty p_n \as^{n+1}
\eeq
we can define a more convergent series
\beq
\hat P(b) = \sum_{n=0}^\infty \frac{p_n}{n!} b^n\;.
\eeq
If this series converges we can define $P(\as)$ by
\beq
P(\as) = \int_0^\infty db\,e^{-b/\as}\,\hat P(b)
\eeq
even when the original series is divergent. Then the
series is said to be {\em Borel summable}. Probably the
QCD perturbation series is not fully Borel summable
[\ref{mueller}], but we may still hope to resum important
sets of contributions in this way.

The growth of QCD matrix elements in the soft region is thought
to give rise to a singularity (an {\em infrared renormalon}) in
the $b$-plane at the position $b = 2/b_0$
($b_0=[11C_A-2\nlf]/12\pi$, called $b$ in earlier
sections), generating a power correction proportional to
\beq
\exp\left[-\frac{2}{b_0\as(Q^2)}\right]
\sim \frac{\Lambda^4}{Q^4}\;.
\eeq
The existence of a renormalon is supposed to
indicate that the full QCD prediction would exhibit the same
power correction. In this language, the appearance of $1/Q$ corrections
to event shapes would be associated with a new infrared renormalon at
$1/2b_0$ in the Borel transforms of these quantities.

In a recent paper [\ref{brw}], an idea about renormalons due to
Bigi, Shifman, Uraltsev and Vainshtein [\ref{BSUV}] was applied to
\ee\ event shapes. Those authors suggested that there is a simple
correspondence between infrared renormalon positions and the power corrections
to fixed-order perturbative predictions evaluated with an infrared cutoff.
In \ee$\to$ hadrons, to first order in $\as$ all diagrams are QED-like and
a suitable cutoff can be imposed by introducing a small mass $\mu$
(in principle much greater than $\Lambda$) into the gluon
propagator.\footnote{Recall that in QED a photon mass can be
introduced in this way without violating the Ward identities
associated with current conservation; see Ref.~[\ref{ItZ}], p.\ 136.}
According to Ref.~[\ref{BSUV}], the first-order perturbative total
cross section with such a cutoff, normalized to the  Born value,
takes the form
\beq\label{Rp}
R_p = 1+\as/\pi - D\,\as\,\mu^4/Q^4 + \ldots
\eeq
where $D$ is a constant and the ellipsis represents non-leading power
corrections.  The dependence on $\mu$ must cancel between
this expression and the soft contribution, which builds the renormalon.
Thus the leading renormalon has to occur at $2/b_0$, as stated above,
and we expect a non-perturbative contribution of the form
\beq\label{Rnp}
R_{np} = [C\,\Lambda^4 + D\,\as(\mu)\,\mu^4]/Q^4
\eeq
where $C$ is a constant. The $\mu$-dependence appears as an arbitrariness
in the part of the correction that we attribute to the renormalon and
the part that is generated in fixed order.

In the case of event shapes, however, one finds that the
introduction of a gluon mass leads to corrections of order
$\as\mu/Q$, which cancel in the total cross section. That is, for a
generic (infrared safe) event shape $S$ one has in first order
\beq\label{Sp}
S_p = A_S\,\as - D_S\,\as\,\mu/Q + \ldots
\eeq
instead of a result of the form (\ref{Rp}).
The coefficients $D_S$ are easily computed; to this order they
arise entirely from the reduction of phase space for real gluon
emission. The values obtained in Ref.~[\ref{brw}] for some representative
quantities are listed, together with the leading coefficients $A_S$,
in Table~\ref{hadro}. Here $T$ is the thrust, $C$ is the $C$-parameter,
and $\sigma_L$ is the longitudinal cross section, as defined
in earlier sections.

\begin{table}
\renewcommand{\arraystretch}{1.5}
\begin{center}
\begin{tabular}{|c|c|c|c|} \hline
      $S$   & $A_S$   & $D_S$ & $C_S\,\Lambda$ \\ \hline\hline
$\VEV{1-T}$ & $0.335$ & $\rat{16}{3\pi} = 1.7$  & $\sim 1.0$ GeV
\\ \hline
$\VEV{C}$   & $1.375$ & 8     & $\sim 5.0$ GeV
\\ \hline
$\sigma_L$  & $1/\pi$ &$4/3$ & $\sim 0.8$ GeV
\\ \hline
\end{tabular}
\end{center}
\caption{\label{hadro}
 Coefficients of terms in Eqs.~(\ref{Sp}) and (\ref{Snp}). }
\end{table}

{}From Eq.~(\ref{Sp}) we expect that a new infrared renormalon
at $1/2b_0$ is present in event shapes, leading to
a non-perturbative contribution
\beq\label{Snp}
S_{np} = [C_S\,\Lambda + D_S\,\as(\mu)\,\mu]/Q \; ,
\eeq
whose dependence on the arbitrary cutoff $\mu$ cancels
against that of the perturbative part, leaving
a cutoff-independent power correction $C_S\,\Lambda/Q$.
The observed values of these corrections, given by Eqs.~(\ref{thad}),
(\ref{chad}) and (\ref{lhad}), are also shown in Table~1.

It is remarkable that the observed $1/Q$ corrections are,
within the uncertainties, proportional to the perturbative
coefficients $D_S$, suggesting that Eq.~(\ref{Snp}) takes the
general form
\beq\label{Snp2}
S_{np} = C_S\frac{\Lambda}{Q}\,\left[1
           + d\,\as(\mu) \frac{\mu}{\Lambda}\right]
\eeq
where $d$ is a constant.  In fact, for the quantities shown, the
cutoff dependence is proportional to the hadronization
correction computed using the simple `tube' model introduced
in Sect.~\ref{sec_mod}. Thus one could argue that the
success of that model is due to some underlying
connection with the renormalon contribution. Similarly the
success of the Monte Carlo hadronization models could be
a reflection of the fact that introducing an infrared
cutoff on the parton shower (called $t_0$ in Sect.~\ref{sec_mod})
is sufficient to reproduce the systematics of the leading
power corrections to event shapes.

It would be interesting to apply the above approach to
a wide variety of quantities, and to try to construct event
shapes from which $1/Q$ corrections are absent. From Table~1
we see that the combination $T+2C/3\pi$ might be of
this type. It would be desirable to extend the treatment
to higher orders in perturbation theory, but it is difficult
to see how this can be done in a gauge-invariant way.\footnote{The
cutoff procedure used in the Monte Carlo models corresponds
implicitly to the choice of a particular axial gauge.}

\section{Conclusions}
\label{sec_conc}
One firm conclusion that can be drawn about hadronization
is that we are still far from understanding it.
More experimental information is needed:
it would be particularly valuable to have data on
power corrections for a wide range of \ee\ observables,
along the lines of Fig.~\ref{meant} for thrust, with proper
analysis of errors on the fitted powers of $Q$ and their coefficients.
Analyses of hadronization corrections to event shape distributions
in terms of {\em smearing} rather than correction factors
would be useful: how do $\sub{\VEV{\delta T}}{had}$ and
$\sub{\VEV{(\delta T)^2-\VEV{\delta T}^2}}{had}$
depend on $T$ and $Q$? Some information on $Q$-dependence
will be obtained from LEP2, but it would be sensible
and profitable to go back and re-analyse as much
lower-energy data as possible in new ways.

The motivation for further experimental and theoretical
study of hadronization is twofold. First, after a long period
of inactivity, apart from some model building, there has
recently been a revival of theoretical interest in
power corrections to perturbative predictions in general.
The interest stems mainly from new ideas about the behaviour of
perturbation theory at high orders, and about the relationship
between perturbative and non-perturbative effects.  Thus there
are new theoretical conjectures to be tested experimentally.

The second motivation is more pragmatic: $\ho 3$ calculations
of more and more observables are steadily becoming available.
To use the extra power of these predictions to measure
$\as$ we need better control of power corrections.  In
particular, it would be useful if $1/Q$ corrections could
be calculated, or if combinations of observables could
be found from which they are absent. As an example, it
was conjectured in Sect.~\ref{sec_pow} that this might
be the case for the quantity $T+2C/3\pi$.

\section*{Acknowledgements}
It is a great pleasure to thank M.\ Locher and all those involved
in organizing the Zuoz Summer School for their warm hospitality.
I am indebted to many people for sharing their illuminating, if
sometimes conflicting, insights into hadronization over the years,
especially my collaborators S.\ Catani, Yu.L.\ Dokshitzer,
V.A.\ Khoze, I.G.\ Knowles, G.\ Marchesini, P.\ Nason and
M.H.\ Seymour.

\par \vskip 15mm
\noindent{\large \bf References}
\begin{enumerate}
\item  \label{QCD20}
       `QCD 20 Years Later', ed.\ P.M.\ Zerwas and H.A.\ Kastrup,
       World Scientific, Singapore (1993).
\item  \label{WebGla}
       B.R.\ Webber, Cambridge preprint Cavendish--HEP--94/15, to
       appear in Proc.\ 27th Int.\ Conf.\ on High Energy Phys.,
       Glasgow, Scotland, 1994.
\item  \label{mueller}
       A.H.\ Mueller, in Ref.~[\ref{QCD20}].
\item  \label{bwarns}
       B.R.\ Webber, \ar{36}{253}{86}.
\item  \label{LPHD}
       Ya.I.\ Azimov, Yu.L.\ Dokshitzer, V.A.\ Khoze and S.I.\ Troyan,
       \pl{B165}{147}{85}; \zp{C27}{65}{85}.
\item  \label{Fey}
       R.P.\ Feynman, `Photon Hadron Interactions', W.A. Benjamin,
       New York (1972).
\item  \label{FF}
       R.D.\ Field and R.P.\ Feynman, \pr{D15}{2590}{77}, \np{B138}{1}{78}.
\item  \label{HOSWZ}
       P.\ Hoyer et al., \np{B161}{349}{79}.
\item  \label{APKW}
       A.\ Ali et al., \pl{B93}{155}{80}.
\item  \label{AKKW}
       A.\ Ali et al., \np{B168}{409}{80}.
\item  \label{AM}
       X.\ Artru and G.\ Mennessier, \np{B70}{93}{74}.
\item  \label{Bow}
       M.G.\ Bowler, \zp{C11}{169}{81}.
\item  \label{AGS}
       B.\ Andersson, G.\ Gustafson
       and B.\ S\"oderberg, \zp{C20}{317}{83},
       \np{B264}{29}{86}.
\item  \label{AGIS}
       B.\ Andersson, G.\ Gustafson, G.\ Ingelman
       and T.\ Sj\"ostrand, \prep{97}{33}{83}.
\item  \label{Sjo}
       T.\ Sj\"{o}strand, \np{B248}{469}{84}
\item  \label{Bar}
       JADE Collaboration, W.\ Bartel et al., \pl{B157}{340}{85}; \\
       TPC Collaboration, H.\ Aihara et al., \zp{C28}{31}{85}; \\
       TASSO Collaboration, M.\ Althoff et al., \zp{C29}{29}{85}.
\item  \label{dipole}
       G.\ Gustafson, \pl{B175}{453}{86};
       G.\ Gustafson and U.\ Pettersson, \np{B306}{746}{88}.
\item  \label{prec}
       D.\ Amati and G.\ Veneziano, \pl{B83}{87}{79};
       A.\ Bassetto, M.\ Ciafaloni and G.\ Marchesini, \pl{B83}{207}{79};
       G.\ Marchesini, L.\ Trentadue and G.\ Veneziano, \np{B181}{335}{80}.
\item  \label{FW}
       R.D.\ Field and S.\ Wolfram, \np{B213}{65}{83}.
\item  \label{clus}
       B.R.\ Webber, \np{B238}{492}{84}.
\item  \label{marweb}
       G.\ Marchesini and B.R.\ Webber, \np{B238}{1}{84}.
\item  \label{DKT}
       Yu.L.\ Dokshitzer, V.A.\ Khoze and S.I.\ Troyan,
       in `Perturbative Quantum Chromodynamics', ed.\ A.H.\ Mueller,
       World Scientific, Singapore (1989).
\item  \label{book}
       Yu.L.\ Dokshitzer, V.A.\ Khoze, A.H.\ Mueller and S.I.\ Troyan,
       `Basics of Perturbative QCD', Editions Frontieres, Paris (1991).
\item  \label{Got}
       T.D.\ Gottschalk, \np{B214}{201}{83}.
\item  \label{PP}
       F.E.\ Paige and S.D.\ Protopopescu, in `Observable Standard
       Model Physics at the SSC: Monte Carlo Simulation and Detector
       Capabilities', Proc.\ UCLA Workshop,
       ed.\ H.-U. Bengtsson et al., World Scientific, Singapore (1986).
\item  \label{Odo}
       R.\ Odorico, \cpc{32}{139}{84}; \cpc{59}{527}{90}.
\item  \label{Fie}
       R.D.\ Field, \np{B264}{687}{86}.
\item  \label{JS}
       T.\ Sj\"{o}strand, \cpc{39}{347}{86};
       M.\ Bengtsson and T.\ Sj\"{o}strand, \cpc{43}{367}{87}.
\item  \label{LEPMC}
       ALEPH Collaboration, D.\ Buskulic et al., \zp{C55}{209}{92}; \\
       L3 Collaboration, B.\ Adeva et al., \zp{C55}{39}{92}; \\
       OPAL Collaboration, R.\ Akers et al., \zp{C63}{181}{94}.
\item  \label{Pyth}
       H.-U.\ Bengtsson and T.\ Sj\"{o}strand, \cpc{46}{43}{87};
       T.\ Sj\"{o}strand, \cern{6488/92}.
\item  \label{Lept}
       G.\ Ingelman, in Proc.\ Workshop on Physics at HERA,
       ed.\ W.\ Buchm\"uller and G.\ Ingelman, DESY, Hamburg (1991).
\item  \label{Ariadne}
       L.\ L\"onnblad, \cpc{71}{15}{92}.
\item  \label{HW}
       G.\ Marchesini and B.R.\ Webber, \np{B310}{461}{88};
       G.\ Marchesini, B.R.\ Webber, G.\ Abbiendi, I.G.\ Knowles,
       M.H.\ Seymour and L.\ Stanco, \cpc{67}{465}{92}.
\item \label{lepevgen}
       T.\ Sj\"ostrand et al., in `Z Physics at LEP 1',
       CERN Yellow Book 89-08, vol.~3, p.~143.
\item  \label{defthrust}
       S.~Brandt, Ch.~Peyrou, R.~Sosnowski and A.~Wroblewski,
       \pl{B12}{57}{64};
       E.~ Farhi, \prl{39}{1587}{77}.
\item  \label{ERT}
       R.K.~Ellis, D.A.~Ross and A.E.~Terrano, \np{B178}{421}{81}.
\item \label{factn}
       J.C.\ Collins, D.E.\ Soper and G.\ Sterman,
       in `Perturbative Quantum Chromodynamics', ed.\ A.H.\ Mueller,
       World Scientific, Singapore (1989).
\item \label{lepqcd}
       Z.\ Kunszt, P.\ Nason, G.\ Marchesini and B.R.\ Webber, in
       `Z Physics at LEP 1', CERN Yellow Book 89-08, vol.~1, p.~373,
       and references therein.
\item  \label{AlP}
       G.~Altarelli and G.~Parisi, \np{B126}{298}{77};
       V.N.\ Gribov and L.N.\ Lipatov, \sj{15}{78}{72};
       Yu.L.\ Dokshitzer, \JETP{46}{641}{77}.
\item\label{NW}
       P.\ Nason and B.R.\ Webber, \np{421}{473}{94}.
\item\label{scaALEPH}
       G.\ Cowan, to appear in Proc.\ 27th Int.\ Conf.\ on
       High Energy Phys., Glasgow, Scotland, 1994; ALEPH Collaboration,
       paper submitted to Glasgow Conference, 1994.
\item\label{BalitskyBraun}
       I.I.\ Balitsky and V.M.\ Braun, \np{361}{93}{91}.
\item\label{angord}
       A.H.\ Mueller, \pl{B104}{161}{81};
       \ B.I.\ Ermolaev and V.S.\ Fadin, \Jlet{33}{285}{81}.
\item  \label{Mueller}
       A.H.\ Mueller,  \np{B213}{85}{83}; Erratum quoted in
       \np{B241}{141}{84}; \np{B228}{351}{83}.
\item  \label{mulex}
       TASSO Collaboration, W.\ Braunschweig et al., \zp{C45}{193}{89}; \\
       ALEPH Collaboration, D.\ Decamp et al., \pl{B273}{181}{91}; \\
       DELPHI Collaboration, P.\ Abreu et al., \zp{C50}{185}{91}.
\item  \label{FongWeb}
       C.P.\ Fong and B.R.\ Webber, \np{B355}{54}{92}.
\item  \label{humpex}
       TASSO Collaboration, W.\ Braunschweig et al., \zp{C47}{187}{90}; \\
       OPAL Collaboration, D.\ Akrawy et al., \pl{B247}{617}{90}.
\item\label{longOPAL}
       OPAL Collaboration, paper submitted to Glasgow Conference, 1994.
\item\label{coeffns}
       G.\ Curci, W.\ Furmanski and R.\ Petronzio, \np{B175}{80}{27};
       \ W.\ Furmanski and R.\ Petronzio, \pl{B97}{80}{437};
       \ E.G.\ Floratos, C.\ Kounnas and R.\ Lacaze, \np{B192}{81}{417}.
\item\label{torbjorn}
       T.\ Sj\"ostrand, private communication.
\item  \label{zeus}
       ZEUS Collaboration, private communication; preliminary
       results presented in Ref.~[\ref{WebGla}].
\item  \label{SVZ}
       M.\ Shifman, A.\ Vainshtein and V.\ Zakharov,
       \np{B147}{385, 448, 519}{79}.
\item  \label{ConSt}
       H.\ Contopanagos and G.\ Sterman, \np{B419}{77}{94}.
\item  \label{MW}
       A.V.\ Manohar and M.B.\ Wise, Univ.\ of California at San Diego
       preprint UCSD/PTH 94-11.
\item\label{brw}
       B.R.\ Webber, \pl{339}{148}{94}.
\item\label{BSUV}
       I.I.\ Bigi, M.A.\ Shifman, N.G.\ Uraltsev and A.I.\ Vainshtein,
       Univ.\ of Minnesota preprint TPI-MINN-94/4-T (CERN-TH.7171/94).
\item \label{ItZ}
       C.~Itzykson and J.B.~Zuber, `Quantum Field Theory',
       McGraw Hill, New York (1980).
\end{enumerate}
\end{document}